\input harvmac

\def\a{\alpha}

\def \na {\nabla}

\def \te {\textstyle}

\def \V {{\cal V}}

\def \te {\textstyle }

\def \V {{\cal V}}

\def \del {\partial}

\def \na {\nabla}

\def \ha{{\textstyle{1\over 2}}}

\def \yg {g_{_{\rm YM}}}

\def \na {\nabla }

\def \a {\alpha}
\def \b {\beta}
\def \ov {\over}
 
\def\r {\rho}

\def \m {\mu}

\def \gym { g^2_{_{\rm YM}} } 
\def \l {\lambda}

\def \P {\Phi}

\def \inv {^{-1}}
\def \ov {\over }
\def \four{{\textstyle{1\over 4}}}
\def \fourth{{{1\over 4}}}

\def \lr { \lref}

\def \four{{\textstyle{1\over 4}}}
\def \fourth{{{1\over 4}}}

\def \del {\partial}

\def \inv  {^{-1}}
 
\def\a{\alpha} 
\def \b {\beta}
 
\def \ov {\over}

\def \half {{1 \ov 2}} 

\def\np {  {\it Nucl. Phys.} }

\lr\KT{I.R. Klebanov and A.A. Tseytlin, ``D-Branes and
Dual Gauge Theories in Type 0 Strings,'' {\tt hep-th/9811035}. }

\lr \DH { L. Dixon and J. Harvey, ``String theories in ten dimensions without space-time supersymmetry",  
{\it Nucl. Phys.} {\bf B274} (1986) 93;
 N. Seiberg and E. Witten,
``Spin structures in string theory", 
{\it Nucl. Phys.} {\bf B276} (1986) 272;
C. Thorn, unpublished.}

\lr\JM {J. Minahan, ``Glueball Mass Spectra and Other Issues for 
Supergravity Duals of QCD Models,'' {\tt hep-th/9811156}.  }

\lr \ferr{G. Ferretti and D. Martelli, ``On the construction of gauge theories from non critical type 0 strings,"
{\tt hep-th/9811208}. }

\lr\Malda {J. Maldacena, ``Wilson loops in large $N$
 field theories," {\it Phys. Rev. Lett.} {\bf 80} (1998) 4859, 
 {\tt hep-th/9803002};
S.-J. Rey and J. Yee, ``Macroscopic strings as heavy quarks in large $N$ gauge theory and anti-de Sitter supergravity", 
{\tt hep-th/9803001}.}

\lr\Sasha{A.M. Polyakov, ``String theory and quark confinement,''
{\it Nucl. Phys. B (Proc. Suppl.)} {\bf 68} (1998) 1, {{\tt hep-th/9711002}}. }

\lr\berg{O. Bergman and M. Gaberdiel, ``A Non-supersymmetric Open
String Theory and S-Duality,'' \np {\bf B499} (1997) 183,
{\tt hep-th/9701137}. }


\lr\GA{ M.R. Garousi, 
``String Scattering from D-branes in Type 0 Theories'',
{{\tt hep-th/9901085}}.
}

\lr  \kleb{
I.R. Klebanov, ``World volume approach to absorption by nondilatonic branes,''
  {\it Nucl. Phys.} {\bf B496} (1997) 231,
  {{\tt hep-th/9702076}}; 
S.S. Gubser, I.R. Klebanov, and A.A. Tseytlin, ``String theory and classical
 absorption by three-branes,'' {\it Nucl. Phys.} {\bf B499} (1997) 217,
  {{\tt hep-th/9703040}}.}

\lr  \gkThree{
S.S. Gubser and I.R. Klebanov, ``Absorption by branes and Schwinger terms in
  the world volume theory,'' {\it Phys. Lett.} {\bf B413} (1997) 41,
  {{\tt hep-th/9708005}}.}

\lr  \jthroat{
J.~Maldacena, ``The Large N limit of superconformal field theories and
  supergravity,'' {\it Adv. Theor. Math. Phys.} {\bf 2} (1998) 231, 
{{\tt
  hep-th/9711200}}.}

\lr  \US{
S.S. Gubser, I.R. Klebanov, and A.M. Polyakov, ``Gauge theory correlators
  from noncritical string theory,'' {\it Phys. Lett.} {\bf B428} (1998)
105,
  {{\tt hep-th/9802109}}.}

\lr  \EW{
E.~Witten, ``Anti-de Sitter space and holography,''
 {\it Adv. Theor. Math. Phys.} {\bf 2} (1998) 253, 
 {{\tt hep-th/9802150}}.}

\lr\GWP{D.J. Gross and F. Wilczek, {\it Phys. Rev. Lett.} {\bf 30}
(1973) 1343; H.D. Politzer, {\it Phys. Rev. Lett.} {\bf 30} (1973)
1346.}

\lr\GW{D.J. Gross and F. Wilczek, 
``Asymptotically free gauge theories 2," {\it Phys. Rev.}  {\bf D9} (1974) 
980. }

\lr  \AP{
A.M. Polyakov, ``The Wall of the Cave,''
{\tt hep-th/9809057.}}

\lr  \brane{
J.~Polchinski, ``Dirichlet Branes and Ramond-Ramond charges,'' {\it Phys. Rev.
  Lett.} {\bf 75} (1995) 4724,
{{\tt hep-th/9510017}}. }

\lr  \Jbook{
J. Polchinski, ``String Theory,'' vol. 2, Cambridge University Press,
1998.}

      \lr\TASI{
J. Polchinski, ``TASI Lectures on D-Branes,''
{\tt hep-th/9611050.}  }

\lr  \Witten{
E.~Witten, ``Bound states of strings and p-branes,'' {\it Nucl. Phys.} {\bf
  B460} (1996) 335, {{\tt
  hep-th/9510135}}.  }

\lr  \hsdgt{
G.T. Horowitz and A.~Strominger, ``Black strings and p-branes,'' {\it Nucl.
  Phys.} {\bf B360} (1991) 197;
M.J. Duff  and J.X. Lu, 
``The selfdual  type IIB  superthreebrane,"
{\it Phys. Lett.}  {\bf B273} (1991)  409;
 G.W. Gibbons and  P.K. Townsend,
``Vacuum interpolation in supergravity via super p-branes",
 {\it Phys. Rev. Lett.} {\bf 71} (1993) 3754, 
{\tt hep-th/9307049}.}

\lr\KW{I.R. Klebanov and E. Witten, ``Superconformal field theory on
threebranes at a Calabi-Yau singularity,'' {\tt hep-th/9807080};
S.S. Gubser, ``Einstein manifolds and conformal field theories,''
{\tt hep-th/9807164}.}
\lr\GNS{
S.S. Gubser, N. Nekrasov and S. Shatashvili,
``Generalized conifolds and four dimensional ${\cal N}=1$
superconformal theories,''
{\tt hep-th/9811230}.
}

\lr\Gir{L. Girardello, M. Petrini, M. Porrati, and  A. Zaffaroni, 
``Novel Local CFT and Exact Results on Perturbations of N=4
   Super Yang Mills from AdS Dynamics", 
{\tt hep-th/9810126};
J. Distler and F. Zamora, 
``Non-Supersymmetric Conformal Field Theories from Stable
   Anti-de Sitter Spaces", 
{\tt hep-th/9810206}.}

\lref \tat {A.A. Tseytlin, 
``On the tachyonic terms in the string effective action", 
{\it Phys Lett.}  {\bf B264} (1991) 311.}
\lref \bank{T. Banks, ``The tachyon potential in string theory",
{\it Nucl. Phys.} {\bf B361} (1991) 166.}

  \lr\bisa{
M. Bianchi and  A. Sagnotti,
``On the Systematics of Open String Theories",
{\it Phys. Lett.} {\bf B247}  (1990) 517.}
\lr\sagn{A. Sagnotti, ``Some Properties of Open - String Theories", 
{\tt hep-th/9509080}; ``Surprises in Open-String Perturbation Theory", 
 {\it Nucl. Phys. Proc. Suppl.} {\bf  B56}  (1997) 332, 
{\tt hep-th/9702093}; 
C. Angelantonj, ``Nontachyonic Open Descendants of the 0B String Theory",
{\tt  hep-th/9810214}. 
}
\lr\mig{A.A. Migdal, ``Hidden Symmetries of Large N QCD,"
{\it Prog. Theor. Phys. Suppl.} {\bf 131} (1998) 269, {\tt hep-th/9610126}.}
  \lr\alva{E. Alvarez, C. Gomez and T. Ortin,
``String representation of Wilson loops", 
{\tt hep-th/9806075}. }

 \lr\jones{D.R.T. Jones,
``Asymptotic behavior  of supersymmetric Yang-Mills theories
in the two-loop approximation," {\it Nucl. Phys.} {\bf  B87}
 (1975) 127;
M.E. Machacek and M.T. Vaughn, 
``Two-loop renormalization group equations in a general 
quantum field theory I: Wave function renormalization,"
{\it Nucl. Phys.} {\bf B222} (1983) 83.}

\lr\baz{T. Banks and A. Zaks,  ``On the
phase structure of vector-like gauge theories with massless
fermions,"
 {\it Nucl. Phys.} {\bf B196} (1982)  189. } 

\lr\gkt{T. Banks and M.B. Green,
``Nonperturbative effects in $AdS_5\times S^5$  
string theory and $d = 4$ SUSY
Yang-Mills," {\it J. High Energy Phys.} 05 (1998) 002, 
{\tt hep-th/9804170}; 
S.S. Gubser, I.R. Klebanov and  A.A. Tseytlin, 
``Coupling constant dependence in the thermodynamics of 
$N=4$ supersymmetric Yang-Mills theory", 
{\it Nucl. Phys.} {\bf B534} (1998) 202, 
{\tt hep-th/9805156}.}

\lr\frat{E.S. Fradkin and A.A. Tseytlin, 
``Quantum properties of higher dimensional and 
dimensionally reduced supersymmetric theories,"
{\it Nucl. Phys.} {\bf  B227}  (1983) 252.}

\lr\wein{S. Weiberg, { The Quantum Theory of Fields},
vol. 2 (Cambridge Univ. Press, 1996).}

\lr \sc {Y. Schr\"oder, ``The static potential in QCD to two loops", 
{\tt hep-ph/9812205}.
}


\baselineskip8pt
\Title{\vbox
{\baselineskip 6pt
{\hbox {PUPT-1825}}{\hbox{Imperial/TP/98-99/20 }}
{\hbox{hep-th/9812089}} 
{\hbox{   }}
}}
{\vbox{\vskip -30 true pt
\centerline {Asymptotic Freedom and Infrared Behavior }
\medskip
\centerline {in the Type 0 String Approach to Gauge Theory}
\medskip
\vskip4pt }}
\vskip -20 true pt 
\centerline{ Igor R. Klebanov}
\smallskip\smallskip
\centerline{Joseph Henry Laboratories, Princeton University, 
Princeton, New Jersey 08544, USA}
\bigskip
\centerline  {Arkady A. Tseytlin\footnote{$^{\dagger}$}{\baselineskip8pt
Also at  Lebedev  Physics
Institute, Moscow.} }
\smallskip\smallskip
\centerline {  Blackett Laboratory, Imperial College,  London SW7 2BZ, U.K.} 

\bigskip\bigskip
\centerline {\bf Abstract}
\baselineskip10pt
\noindent
\medskip
In a recent paper
we considered the type $0$ string theories, obtained from 
the ten-dimensional closed NSR string  by a GSO
projection which excludes space-time fermions,
and studied the low-energy dynamics of $N$ coincident D-branes.
This led us to conjecture that the four-dimensional 
$SU(N)$ gauge theory coupled
to 6 adjoint massless scalars is dual to a background of type 0
theory carrying $N$ units of R-R 5-form flux
and involving a tachyon condensate. The tachyon background
leads to a ``soft breaking'' of conformal invariance, and we derived the
corresponding renormalization group equation.
Minahan has subsequently found its asymptotic
solution for weak coupling and showed that 
the coupling exhibits logarithmic flow, as expected from the asymptotic 
freedom of the dual gauge theory. We study this solution in more
detail and identify the effect of the 2-loop beta function.
We also demonstrate the existence of a fixed point at infinite coupling.
Just like the fixed point at zero coupling, it is characterized by
the $AdS_5\times S^5$ Einstein frame metric.
We argue that there is a RG trajectory extending all the way from the
zero coupling fixed point in the UV to the infinite coupling fixed
point in the IR.

\bigskip
 
\Date {December 1998}

\noblackbox \baselineskip 15pt plus 2pt minus 2pt 

\newsec{Introduction}

In a recent paper \refs{\KT} we conjectured that 
$3+1$ dimensional $SU(N)$ gauge theory
coupled to 6 adjoint massless scalar fields is dual to a certain
background of type 0 string theory \DH\ involving a non-vanishing 
tachyon field.
This work was inspired by the recently discovered relations between
type II strings and superconformal gauge theories on $N$ coincident
D3-branes \refs{\kleb,\gkThree,\jthroat,\US,\EW}, as well as by
Polyakov's suggestion \refs{\AP} (building on his earlier work \Sasha) 
that the type 0 string theory in dimensions $D \leq 10$ is a natural
setting for extending this duality to non-supersymmetric non-conformal
gauge theories.\foot{Investigation  of some aspects 
of Polyakov's proposal in the non-critical case appeared
recently in \ferr.} 

The type 0 string has world sheet supersymmetry,\foot{
Possible relevance of world sheet supersymmetry to string description of
gauge theories was also advocated in \refs{\Sasha,\mig,\alva}.}
but the GSO projection is non-chiral and breaks the space-time
supersymmetry. Following the notation of \refs{\Jbook}, in
$D=10$ the spectra of the type 0A and type 0B theories are:

\noindent
$
\ \ \ \ \ \ \ \ {\rm type} \ 0A : \ \ \ 
(NS-,NS-)\oplus (NS+,NS+)   \oplus(R+,R-) \oplus(R-,R+)\ ,$

\noindent  $
\ \ \ \ \ \  \ \ {\rm type} \ 0B: \  \ \ (NS-,NS-)\oplus(NS+,NS+) 
\oplus(R+,R+) \oplus(R-,R-)\ .$

\noindent
Both of these theories have no fermions in their spectra
but produce modular invariant
partition functions \refs{\DH,\Jbook}.
The massless bosonic fields are  as in the corresponding 
type II theory (A or B), 
but with the  doubled set of the Ramond-Ramond (R-R)  fields. 
The type 0 theory also contains a tachyon from the $(NS-,NS-)$ sector,
which is why it has not received much attention thus far.
In \refs{\AP,\KT} it was suggested, however, that the presence of the
tachyon does not spoil its application to large $N$ gauge theories.
A well-established route towards gauge theory is via the 
D-branes \refs{\brane,\Witten},
which were first considered in the type 0 context in \berg.
Large $N$ gauge theories, which are constructed on $N$ coincident D-branes
of type 0 theory, may be shown to contain no open string
tachyons \refs{\KT,\AP}.\foot{The open  string
 descendants of type 0B theory
were originally constructed  by orientifold projection 
in \bisa. They, in general, have open-string tachyons in their spectra. 
A non-tachyonic model  
(which is anomaly-free and contains chiral fermions in its 
open-string  spectrum) was  found 
by a special Klein-bottle projection of
the 0B theory in \sagn. We will be concerned with a different tachyon-free
theory which occurs on parallel like-charged D-branes.} 
Furthermore, the dual type 0 background necessarily includes $N$
units of R-R flux which has a stabilizing effect on the bulk tachyon
\KT.
In fact, in \KT\ the presence of a tachyon background was turned into
an advantage because it gives rise to the renormalization group (RG) flow. 

In \KT\ the $3+1$ dimensional $SU(N)$ theory coupled to 6 adjoint massless
scalars was constructed as the low-energy description of $N$
coincident electric D3-branes. The conjectured dual type 0
background thus carries $N$ units of electric 5-form flux.
In the Einstein frame the dilaton decouples from the $(F_5)^2$ terms in
the effective action, and the only source for it originates
from the tachyon mass term,
\eqn\uiu{ \nabla^2 \P ={\te{ 1\over 8} }  m^2
e^{{1\ov 2}\P} T^2
\ , \ \ \ \ \ \ \ \  m^2 = - { 2 \ov \a'} \ . 
}  
Thus, the tachyon background induces a radial variation of $\Phi$.
Since the radial coordinate is related to the energy scale of
the gauge theory \jthroat, the effective coupling decreases
toward the ultraviolet (UV) \KT, 
in agreement with the expected asymptotic freedom
of the  gauge theory \GWP.

Further progress was recently achieved by Minahan, who found the
asymptotic UV (large radius)
form of the solution to the equations for the type 0
background proposed in \KT. In addition to the ``RG equation''
 \uiu,
one has   
$$ R_{mn} - \ha  g_{mn} R = \four \na_m T \na_n T -
{\te{ 1\over 8} } g_{mn} [(\na T)^2 + m^2 e^{{1\ov 2}\P} T^2] + 
\te{1\over 2} \nabla_m \P\nabla_n \P 
- \four g_{mn} (\na \P)^2
$$ 
\eqn\fii{
  + \ 
{\textstyle{1 \ov 4 \cdot 4!}} f(T) 
(F_{m klpq} F_{n}^{\ klpq}
 - {\textstyle { 1\ov 10}} g_{mn} F_{s klpq} F^{s klpq}) \ , }
\eqn\fiy{ (-\nabla^2 + m^2 e^{{1\ov 2}\P} ) T + 
  {\textstyle{1 \ov 2 \cdot 5!}}  f'(T) F_{s klpq} F^{s klpq} =0 \ , \  }
\eqn\more{ \na_m [ f(T) F^{mnkpq} ] =0 \ ,}
where
$g_{mn}$ is the Einstein frame metric, and
the tachyon--R-R field coupling function is \KT
\eqn\fuu{
f(T) \equiv 1 + T + \ha T^2  \ . 
}
Assuming as in \KT\ that
the tachyon  is  approximately 
localized near  the  extremum of $f(T)$, i.e.  
$T=-1$, Minahan found \JM\ that $g_{mn}$ is
asymptotic to $AdS_5\times S^5$,
while the effective string coupling is
\eqn\minn{ {1\over g_{st}} = e^{-\Phi} \sim  N \ln^2 u 
\ .
}
Here $u$ is the radial coordinate of $AdS_5$
and $N \sim Q$ is the number of R-R flux units. 

The logarithmic flow of the effective coupling is very encouraging.
Furthermore, the calculation \JM\ of the quark-antiquark potential
following the prescription of \Malda\ gives\foot{Note that 
the semiclassical value of the Wilson loop 
is determined by the fundamental string  action,
which contains the  string-frame metric $G_{mn} = e^{\half \P} g_{mn}$.}
\eqn\effpot{ {\cal V }  \approx  - 
{ k_1 \over L \ln {L_0 \ov L}  }
\ ,\qquad\ \ \ \ 
L\ll L_0\ , \ \ \ \ \ \ \   k_1= \left [{4\ov \Gamma(\fourth)}\right ]^4\ , 
}
in agreement with the short-distance behavior of the gauge theory \GWP: 
\eqn\vvv{
{\cal V}\ \sim\  -{ N\gym(L) \ov L} \ , \ \ \ \ \ \ \    N\gym(L) 
\sim { 1 \ov \ln {L_0 \ov L} } \  .}
The logarithmic term in the denominator appears as
follows. In the $AdS_5$ calculation of \Malda\ the
quark-antiquark potential determined from the
Wilson loop factor (i.e. the semiclassical value of the 
fundamental string action)  
is proportional to $\sqrt{g_{st} N}$,
which is constant for ${\cal N}=4$ SYM theory. While the theory
we are studying is non-conformal, $\sqrt{g_{st} N} \sim { 1\ov \ln u}$ varies 
slowly for large $u$.
If we consider a quark and an antiquark separated by distance $L$,
the string connecting them penetrates to $u$ of order $1/L$ due to the
approximate conformal invariance. Introducing the ``QCD scale'' $L_0$,
we have $u\sim L_0/L$, so that the potential is multiplied by the
effective value of $\sqrt{g_{st} N}$, which is $1/\ln (L_0/L)$.

Finding the result \effpot\ from the dual type 0 string description
is striking, but it is important to study various corrections and show
that they do not destroy it. In this paper we consider, in particular, 
the $R^4$
correction to the effective action and show that it does not change
the scaling found from the leading  effective gravity solution, though it does
change the coefficients. This conclusion actually applies to
a whole class of possible $\alpha'$ corrections.
Thus, in order to compare the
beta function coefficients, it seems necessary to know the exact
string $\sigma$-model, but the effective gravity does capture the
physics of the dual gauge theory.

Another important physical question is what happens to the theory in
the infrared (IR) limit. One possibility is that the adjoint scalar fields
become massive, so that the theory is in the same universality class
as the pure glue $SU(N)$ theory. On the dual string side this would
manifest itself in the disappearance of the 5-sphere.
In this paper we explore a different possibility: that in the
IR limit the scalars remain massless and
the 5-sphere remains macroscopic.
We first observe that, as the theory starts flowing from
the UV toward longer distance
scales, the tachyon begins to shift from $-1$ towards $0$. In fact,
we succeed in finding an asymptotic IR (small $u$) solution where the coupling
increases logarithmically, the tachyon approaches zero,
while the Einstein metric is 
$AdS_5\times S^5$. The physical interpretation of this is that the
theory flows towards a fixed point at infinite coupling.
Thus, instead of a confining theory in the infrared, we find a conformally
invariant theory with infinite coupling!

The existence of an IR fixed point in  $SU(N)$ gauge theory
with 6 adjoint scalars  is 
not unexpected:  as we shall discuss below, 
the one-loop coefficient in the 
beta-function is negative  but the two-loop one is positive 
so that, in addition to the UV attractive fixed point at $\gym=0$,
in the two-loop approximation there 
is an IR attractive fixed point \GW\ at $\gym \sim 1/N$. 
However, since in this approximation
the IR fixed point is found to be at rather large `t Hooft coupling,\foot{
This is in contrast to  what happens in theories with 
the number of matter  fields of order $N$,
where a similar IR fixed point is at
 $\gym \sim 1/N^2$ \baz.} 
its position  is  expected  to be  changed  by 
higher-order  corrections. Our dual gravity description suggests 
 that the fixed point is actually  shifted  to
 infinite coupling.

In Section 2 we study the $u \gg 1$  (UV)
 solution in detail and derive
corrections to the asymptotic form of the solution of \JM. We also estimate
the effect of the $\alpha'$ corrections.
In section 3 we discuss the perturbative expression for 
the beta-function in the $SU(N)$ gauge theory  with 6 adjoint 
scalars. We interpret the $\ln\ln u$  corrections to the leading 
UV solution  as corresponding to the 2-loop 
terms in the running gauge coupling constant. 
In section 4 we derive the $u \ll 1$ (IR)
asymptotic solution and discuss its physical interpretation.
Section 5 contains some concluding  remarks.

\newsec{UV asymptotic solution: the asymptotic freedom}
Parametrizing  the  10-d string-frame and Einstein-frame  metric 
as in \KT \ ($\m=0,1,2,3$ are the 4-d indices) 
\eqn\met{
ds^2 = e^{\half \P} ds^2_E\ , \ \ \ \ \ \ \ 
ds^2_E = e^{\half \xi - 5 \eta} d\r^2  + e^{-\half \xi}  dx^\m dx^\m
 + e^{\half \xi -  \eta} d\Omega^2_5 \ , }
the radial effective action corresponding to \uiu--\more\ 
becomes \KT
\eqn\eas{  
S=  \int d\r \bigg[ \ha \dot  \P^2  + \ha \dot \xi^2   
- 5  \dot \eta^2 + \four \dot T^2
   - V(\P,\xi,\eta,T) \bigg] \ , 
}
\eqn\eio{
V =  \ha  T^2  
e^{{1 \ov 2}\P + {1 \ov 2}  \xi   - 5 \eta }
   + 20   e^{-4\eta }   -   Q^2   f^{-1} (T)\  e^{-2\xi}  \ . } 
Here $\a'=1$ 
and  $\P, \xi, \eta$  and $T$ are functions of $\r$.
The constant $Q$ (the R-R charge)  can be absorbed 
into  the 
 redefinition  $\xi\to \xi + \ln Q$ and $\P\to \P - \ln Q$ and may be 
set equal to 1 in intermediate calculations. 
The  resulting set of variational equations,
\eqn\sett{ \ddot \Phi+ {1\over 4} T^2
e^{{1 \ov 2}\P + {1 \ov 2}  \xi   - 5 \eta }=0\ , }
\eqn\sees{ \ddot \xi + {1\over 4} T^2
e^{{1 \ov 2}\P + {1 \ov 2}  \xi   - 5 \eta } + 2 Q^2 f^{-1} (T)
e^{-2\xi}=0\ ,    }
\eqn\sss{ \ddot \eta + 8 e^{-4\eta } + {1\over 4} T^2
e^{{1 \ov 2}\P + {1 \ov 2}  \xi   - 5 \eta }
=0 \ ,}
\eqn\ss{ \ddot T + 2 T e^{{1 \ov 2}\P + {1 \ov 2}  \xi   - 5 \eta }+
2 Q^2 {f'(T)\over f^2(T)} e^{-2\xi}=0  \ ,    }
should be supplemented by the `zero-energy' constraint
\eqn\cons{
\ha \dot  \P^2  + \ha \dot \xi^2   
- 5  \dot \eta^2 + \four \dot T^2
   +  V(\P,\xi,\eta,T) =0 \ , }
which can be  used instead of one of the second-order equations.

Let us recall that
if one ignores the  first term  in \eio, which originates from the
tachyon mass, 
and  takes  the tachyon to be at the extremum of $f(T)$
($T=-1$, \ $f(-1)=\ha$), then 
one finds \KT\ the   electric  analogue 
of the standard  R-R charged 3-brane solution 
($\r =  { e^{2\P_0}  \ov 4r^4} = { 1 \ov u^4}$) 
\eqn\soo{
T=-1\ , \ \ \ \ \ \ \   \P=\P_0 \ , \ \ \ \ \ 
e^\xi  = e^{\P_0}  + 2 Q  \r\ , \ \ \ \ 
e^\eta = 2 \sqrt \r \ . \ }
This becomes  the  $AdS_5 \times S^5$ space 
\eqn\sios{
T=-1\ , \ \ \ \ \  \P=\P_0 \ , \ \ \ \ \ 
\xi  =   \ln (2Q) + \ln  \r  \ , \ \ \ \  \ 
\eta =  \ln 2 +  \ha  \ln  \r \   }
in the  near-horizon ($2 Q  \r \gg e^{\P_0}$) 
 limit, i.e. ($\r= u^{-4}$) 
\eqn\ads{ ds^2_E=  R^2_0  ( {du^2\ov u^2}  + {u^2\ov 2 R^4_0} dx^\m dx^\m 
+ d \Omega^2_5  ) \ , \ \ \ \ \ \ \ \   
 R^2_0 = 2^{-1/2} Q^{1/2}\ .  } 
In \JM\ it was noted that, since $e^\Phi$ becomes small in the
UV region \KT, the tachyon mass term acts as a small perturbation.
We shall indeed confirm below 
that there is a systematic expansion giving
a solution of the full equations \sett\ which is asymptotic to \ads.

Let us study the full set of equations for small $\rho$ by defining
 \eqn\dee{
\rho \equiv  e^{-y} \ll 1 \ , \ \ \ \ \  \ \ \ \ \ y \gg 1 \  . }
By direct inspection of the system 
of equations  we find the following  asymptotic 
solution\foot{We thank J. Minahan for pointing out some wrong
numerical coefficients in the original version of this paper.}
\eqn\sou{
T= -1  +   {8\ov y}
 +  { 4\ov y^2}   ( 39 \ln y  - 20) + O ( { \ln^2 y \ov y^3}) \ , }
\eqn\souu{
\Phi= \ln (2^{15}Q\inv)  -  
2  \ln y  +    {1\ov y}  39\ln y  + O({ \ln y\ov y^2 }) \ , }
\eqn\suu{
\xi= \ln (2 Q)  - y   +    {1\ov y}
 +  {  1\ov 2 y^2}   ( 39 \ln y  - 104) + O ( { \ln^2 y \ov y^3}) \ , }
\eqn\suuu{
\eta= \ln 2  - \half  y   +    {1\ov y}
 +  {  1\ov 2 y^2}   ( 39 \ln y  - 38) + O ( { \ln^2 y \ov y^3}) \ .  }
The leading $O(\ln y)$ term in $\P$, and the $O(y)$ and
$O(1/y)$ terms in $\xi$ and $\eta$ 
were found in \JM.  What we have shown is
that the leading-order solution
of \JM,  which itself reduces to $AdS_5 \times S^5$ 
in the $\rho \to 0$ ($y\to \infty$)  limit, 
can be systematically extended to a solution of the full set 
of equations, including the tachyon one.  The evolution of the 
tachyon (which starts at the  extremum $T=-1$ of $f(T)$ and 
grows at larger $\rho$ towards $T=0$) is crucial for  the 
consistency of the solution.

As a result, the inverse coupling is 
\eqn\coup{
e^{-\P} = 2^{-15} Q \ y^{2  - {39\ov y} + O({ 1\ov y^2 })}=
 2^{-15} Q\ (\ln \r)^{2  + {39\ov \ln \r} + O({ 1\ov \ln^2\r })} \ , }
and the 10-d  Einstein-frame metric \met\ is  (cf. \ads) 
$$
ds^2_E = R^2_0 \bigg[  \  ( 1 - {9\ov 2y}   - 
{351\ov 4y^2} \ln y + ... )   ({1\ov 4}dy)^2 $$
\eqn\meet{
+ \  (1 - { 1\ov 2y} - { 39\ov 4y^2} \ln y + ... )\ {e^{{1\ov 2}y}\ov 2 R^4_0} \ 
dx^\m dx^\m 
+ 
 (1 - { 1\ov 2y} - { 39\ov 4y^2} \ln y + ... ) d \Omega^2_5  \bigg] \ , }
where
$$\ \ \ \ \   R^2_0 = 2^{-1/2} Q^{1/2}\  . $$
Noting that $y = 4 \ln u$, we see that
this metric starts as $AdS_5 \times S^5$ \ads\  
at $y =\infty$  and  becomes a 10-d space with negative Ricci scalar
at smaller $y$ (bigger $\r$).  The corrections
to $\xi$ and $\eta$ cause the effective radius  of $AdS_5$ to become
smaller than that of $S^5$, leading to the 
negative deficit in the total curvature (cf. \JM). 

While it is satisfying to find the asymptotic freedom from the
large $u$ behavior of the  leading effective gravity solution, 
the most important
question is whether it survives the full string theoretic treatment.
Naively, the string scale corrections should be large because $g_{st} N$
is becoming small for large $u$, but,  remarkably,  
the solution is robust due to
its special structure related to the approximate conformal invariance.
One fact that is crucial for our purposes is that the Einstein metric
is asymptotic to $AdS_5\times S^5$. This geometry is
conformal to flat space, so that the Weyl tensor vanishes
in the large $u$ limit.
Furthermore, both $\Phi$ and $T$ vary slowly for large $u$.

Let us consider, for instance, the leading $\alpha'^3$ correction
to the effective action. 
 In the Einstein frame, the dilaton equation
becomes
\eqn\newdil{ \nabla^2 \P =-{{ 1\over 4\alpha'} } 
e^{{1\ov 2}\P} T^2 + {3\over 16}\a'^3 \zeta(3)  e^{-{3\ov 2}\Phi} W
\ ,
}  
where $W$ is built out of the fourth power of the Weyl
 tensor\foot{As
 explained in \KT\ the  tachyon-independent terms 
in the tree-level
 effective action of 
type 0 string theory are the same as in type II theory.
For a discussion of the $R^4$-correction in a similar 
context in type II theory see \refs{\gkt} and references  there.}
\eqn\weyl{
W =  C^{hmnk} C_{pmnq} C_{h}^{\ rsp} C^{q}_{\ rsk} 
 + \half  C^{hkmn} C_{pqmn} C_h^{\ rsp} C^{q}_{\ rsk}  
 \  . 
}
Evaluating the Weyl tensor on the solution, 
we find that $|C| \sim {1\ov \ln u}$, so that 
\eqn\soyo{ W\sim {1\over \ln^4 u}\ .
}
Since for large $u$  the inverse  coupling  grows as 
  $e^{-\half\Phi} \sim {\ln u}$, we see that the
$\alpha'^3$ correction to the dilaton equation is {\it of the
same order}  as
the leading order gravity contribution. We believe that this is a general
feature of the solution. For instance, other admissible 
higher-order correction terms
which are of the schematic form
$$  e^{{1 \ov 2} \P}   (e^{ -{1\ov 2} \P} C)^n 
\ ,
$$
also scale as $1\ov \ln u$. 
Another class of terms that could
appear on the right-hand side of the Einstein-frame 
dilaton equation is
$e^{-\half (n-1)\Phi} (\nabla^m\P\nabla_m \P)^n$, which
turns out to be subleading (it scales as $(\ln u)^{-n}$).\foot{
Terms involving the Ricci tensor
or $\nabla^2 \P$ can be traded for other terms
by use of 
the equations of motion or, equivalently, by field redefinitions.
For example, $e^{-\half (n-1)\Phi} (\nabla^2\P)^n$
turns out to be
of the same order as $e^{{1\ov 2}\P} T^2$. }

These considerations
give us some confidence that the asymptotic freedom evident in the
leading order gravity approach survives the full string $\sigma$-model
treatment.  As for the string loop corrections, they  
are further suppressed  for $u \gg 1$ 
by powers of $e^{\P}  \sim { 1 \ov Q \ln^2 u}$.

\newsec{Correspondence with the 
  two-loop   RG evolution of gauge  coupling}

We can further ask about the perturbative corrections to the RG
flow. From the expression for the dilaton 
\souu,\coup\ we have ($y= 4 \ln u$, cf. \minn)
\eqn\couyy{   \sqrt{g_{st} N} \sim 
Q^{1/2} e^{\half \P}\sim 
{1\over \ln u -{39\over 8} \ln \ln u + ... }
\ .
}
In calculating the quark-antiquark potential, the relevant value of
$u$ for the location of the string is of order $L_0/L$. Thus,
we estimate the correction to the potential \effpot\ to be
\eqn\neweffpot{ {\cal V } \approx -  { k_1 \over L \ (\ln {L_0\ov L}
-{39\over 8} \ln \ln {L_0\ov L} ) }\ ,\qquad \ \ \ \ 
L\ll L_0\ .
}

A more precise solution for the shape of the string in the metric
\meet, as well as the $\alpha'$ corrections, 
can change the coefficients in \couyy.
Nevertheless, it is  satisfying
that \neweffpot\  does have the same structure 
(including the relative sign of the $\ln (L_0/L)$  and 
the $\ln\ln (L_0/L)$ terms) as in the 
dual gauge theory with the two-loop  term in the 
beta function taken into account!

Let us demonstrate this in detail. The 
gauge theory  in question  can be thought of as
a  reduction of YM theory from 
10 to 4 dimensions (or as a truncation 
of the $\cal N$=4  SYM theory which removes the fermions).
The bare values of the gauge and quartic scalar couplings  
may be different since in the absence  of supersymmetry their 
equality at the tree level does not survive 
renormalization.  The  quartic scalar coupling does not contribute 
to the 2-loop beta-function of the gauge coupling,  so that 
the RG equation for the YM coupling is 
\eqn\rgg{
L { d g_{_{\rm YM}} \ov d L} =    b_1 {\yg^3\ov (4\pi)^2} +  b_2{
 \yg^5 \ov (4\pi)^4 } + ...
\ ,  }
where $L$ is the coordinate scale $L \ll L_0$.
 The solution of this  RG equation is  (see, e.g., \wein)
\eqn\soli{
\gym(L) = { (4\pi)^2 \over 2b_1 \ln {L_0 \ov L} }
 \bigg [ 1 - { b_2\over 2 b_1^2}
 { \ln \ln {L_0 \ov L}  \over  \ln {L_0 \ov L} }\bigg] + ...
= { 8 \pi^2 \over b_1 (  \ln {L_0 \ov L}  +  { b_2\over 2 b_1^2}
  \ln \ln {L_0 \ov L}  ) }  + ...
\ . } 
The  one- and two-loop coefficients $b_1$ and $b_2$  in  a 
gauge theory  with group $G$ coupled
to scalars  transforming in a representation $R$ 
are  \jones\ 
\eqn\twol{
b_1 ={ 11\ov 3}  C_2(G) -   { 1\ov 6 } T_2(R)     \ , }
\eqn\ttww{
b_2 ={ 34\ov 3 } [C_2(G)]^2     - [2C_2(R) +  {1\ov 3} C_2(G)] T_2 (R)
\ , } 
where 
$T_2$ is 
 the Dynkin index of the  representation $R$, 
$\Tr (T_A T_B) = T_2 (R) \delta_{AB}$, 
and $C_2(R)$ is the eigenvalue of the 
quadratic Casimir in this representation. We have 
$ d(G) T_2 (R)  = d(R) C_2 (R)$
where $d(R)$ and $d(G)$ are  the dimensions  of the 
representation and the group.
For the adjoint representation of $SU(N)$
$$ T_2(R) = C_2(R) = C_2(G)=N \ .
$$
For the case of the $SU(N)$ theory coupled to
$N_s$ adjoint scalars the above formulae imply\foot{
Note, in particular,  that the theory obtained by reduction 
of the $D=26$ YM theory ($N_s=22$)  which has a vanishing one-loop 
beta-function \frat, has $b_2 < 0$ and thus is not asymptotically free.}
\eqn\rres{
 b_1 = {1\ov 6} (22-N_s) N\ , \ \ \ \ \ \ \ \ \ 
   b_2 = {1\ov 3}  (34-7N_s) N^2 \ , }
so that for $N_s =6$ we finally have 
\eqn\yut{
 b_1 = {8\ov 3}  N\ , \ \ \ \ \ \ \ b_2 = - {8\ov 3} N^2 
\ . }
Thus, the relative coefficient between the $\ln \ln (L_0/L)$
term and the $\ln (L_0/L)$ is 
$${b_2\ov 2 b_1^2}= -{ 3\ov 16}  \ .  $$
This differs from the $-39/8$ found in the type 0 calculation
\couyy, 
but the sign of the effect is correct.

Since $b_1 > 0$,  this  gauge theory 
has the usual asymptotically free
 UV attractive fixed point at $\yg=0$. Since $b_2 < 0$,
in the 2-loop approximation there is also
an IR attractive fixed point (cf. \GW)  at 
\eqn\fixed{ \gym =  - (4\pi)^2  {b_1 \ov b_2} =
 { (4\pi)^2 \ov  N} \approx { 158 \ov N} \ . }
However, this possible fixed point is located at a rather
large value of the `t Hooft coupling $\l = \gym N$ where the
perturbative analysis is obviously not reliable: 
it  should be shifted by the higher-order 
$\l^n$ corrections in the large $N$ 
perturbation theory for the beta function.
In fact, the discussion of the IR gravity solution in section 4
suggests that this fixed point is shifted to infinite 
coupling.

\newsec{IR asymptotic solution: the fixed point at infinite coupling}

In the previous sections we found the small $\r$ expansion of a
RG trajectory which originates from a UV fixed point at vanishing coupling.
One interesting feature of the trajectory is that $T$ starts increasing
from its critical value $T=-1$ determined from the condition $f'(T)=0$.
The precise form of the trajectory for finite $\rho$ is not known 
analytically, but we can extract some qualitative features from
the RG equations \sett--\ss.\foot{Recently RG flows in gauge theories
were studied from the type IIB perspective in \refs{\KW,\Gir,\GNS}. 
There the context
is somewhat different since the flow connects lines of fixed points.} 

We note that the fields $\Phi$, $\xi$ and $\eta$ have negative
second derivatives. Thus, each of these fields 
may reach a maximum at some value of $\rho$.  
If for $\Phi$ this happens at a finite $\rho$, then we reach
a peculiar conclusion that the coupling is decreasing far in the infrared.
In fact, a different possibility seems to be realized: $\dot \Phi$ is
positive for all $\rho$, asymptotically vanishing as $\rho\rightarrow
\infty$. We have succeeded in constructing the asymptotic form of such
trajectory. It is crucial that, as $\rho\rightarrow \infty$,
$T$ approaches zero so that, as in the UV region, $T^2 e^{\half \Phi}$
becomes small. For this reason the limiting Einstein-frame
metric is again $AdS_5\times S^5$.
 Thus, the theory flows to a conformally invariant
point at infinite coupling.

Indeed, in the region 
 \eqn\larg{\rho \equiv  e^{y} \gg 1 \ , \ \ \  \ \ \ \ \ \ \ \ 
 y \gg 1 \  , }
we find the following  asymptotic  large $\r$ solution 
\eqn\souo{
T=    - {16\ov y}
 -  {  8\ov y^2}   ( 9 \ln y  - 3) + O ( { \ln^2 y \ov y^3}) \ , }
\eqn\souuo{
\Phi= - \half  \ln (2 Q^2)   +  
2  \ln y  -     {1\ov y} 9 \ln y  + O({ \ln y\ov y^2 }) \ , }
\eqn\suuo{
\xi=  \half \ln (2 Q^2)  +  y   +    {9\ov y}
 +   {  9\ov 2 y^2}   ( 9 \ln y  - {20\ov 9}
) + O ( { \ln^2 y \ov y^3}) \ , }
\eqn\suuuo{
\eta= \ln 2  +  \half y   +    {1\ov y}
 +  {  1\ov 2 y^2}   ( 9 \ln y  - 2) + O ( { \ln^2 y \ov y^3}) \ .  }
Its Einstein-frame metric 
is asymptotic to $AdS_5 \times S^5$  at $\rho=\infty$
and evolves into a metric with negative Ricci scalar at smaller $\rho$.
The structure of the asymptotic UV and IR solutions is
similar (notice that the  coordinates  corresponding
to the two regions  are related by
$y \to -y$, cf. \dee--\suuu),  
suggesting that they  can be smoothly connected into the full 
interpolating solution. 
This is indeed supported by
our numerical analysis which shows, in particular,
that  the tachyon  starts at $T=-1$ at  $\r=0$, grows
according to \sou, then enters an oscillating regime and finally relaxes
to zero according to \souo. 

The fact that $T$ goes to zero makes the details of the coupling function
$f(T)$ largely irrelevant in this  IR region, 
implying that the resulting picture is quite robust.
In the IR the dilaton blows up, i.e.
the coupling is 
\eqn\coupa{
e^{\P} =  2^{-1/2} Q^{-1}  y^{2  - {9\ov y} + O({ 1\ov y^2 })} \ , }
and the 10-d  Einstein-frame metric \met\ is   (cf. \meet) 
$$
ds^2_E =  R^2_\infty \bigg[\   ( 1 - {1\ov 2y}   - 
{9\ov 4y^2} \ln y + ... )   ({1\ov 4} dy)^2 $$
\eqn\meta{
+ \
  (1 - { 9\ov 2y} - { 81\ov 4 y^2} \ln y + ... ) 
  {e^{-{1\ov 2}y}\ov 2 R^4_\infty} \ 
 dx^\m dx^\m   + 
 (1 + { 7\ov 2y}  + { 63\ov 4y^2} \ln y + ... ) d \Omega^2_5  \bigg] \ ,  }
where 
$$  R^2_\infty = 2^{-3/4} Q^{1/2}\ .  $$
This metric starts again 
 as $AdS_5 \times S^5$ \ads\  ($y = -4 \ln u$) 
at $y =\infty$  and  becomes a negative curvature 10-d space at smaller
$y$ (bigger $u$).  As in the 
large $u$  region, the  corrections
to $\xi$ and $\eta$ cause the  effective radius  of $AdS_5$  to become 
smaller than that of $S^5$, 
leading to the negative  total  scalar curvature. 

Indeed, the scalar curvature 
in the Einstein frame is 
\eqn\sca{
R =  -{ 5\ov 8  } T^2 e^{\half \P}  + \half  g^{mn} \del_m \P \del_n
\P  \ , } 
where the  first term gives the dominant
$O(1/y)$ contribution. We find that in the 
 UV regime
the $1/y$ corrections to the metric \meet\ give ($\r= e^{-y} $)
\eqn\aqs{ R(\r\to 0)  = -  
{ 80 \ov R^2_0} { 1 \ov  y } +  O({ \ln y\ov y^2 })  \ ,
}
while in the IR regime 
the metric \meta\ gives ($\r= e^y $)
\eqn\aqsnew{R(\r\to \infty)  = -  
{ 80 \ov R^2_\infty} { 1 \ov  y } +  O({ \ln y\ov y^2 })  \ .
 } 
Thus, in both regimes the Ricci scalar
becomes negative away from the critical point.
The absolute value of the 
 curvature is small in both asymptotic regions but 
grows in between.

Note also that for the IR solution all higher-order $\alpha'$ corrections
are suppressed. The string loop corrections may become important right at
the fixed point. Away from the fixed point, $g_{st}$
is of order $1/Q$ and is thus regarded as very small.

\newsec{Discussion}

We have presented new evidence for our conjecture \KT\ that
$SU(N)$ gauge theory coupled to 6 massless adjoint scalars is dual to
type 0 string theory. The type 0 formulation indicates the presence of
two fixed points: one at vanishing coupling and the other at infinite
coupling. The RG flow for weak coupling has a number of features expected
from the asymptotic freedom of the dual gauge theory, including the effect
of the two-loop correction to the beta function. We have further argued
that, as the RG trajectory is continued towards the infrared, the coupling
grows indefinitely, eventually reaching the fixed point at infinite coupling.
We constructed the asymptotic expansion near this IR fixed point, but
it would be of further interest to study the entire trajectory in
detail.

In all our calculations we used a specific form \fuu\ 
of the function $f(T)$
describing the tachyon couplings to the 5-form gauge fields.
One may be concerned that the form of this function depends on the
scheme adopted in the derivation of the effective action (for example,
it may be changed by field redefinitions).
We believe that practically any function $f(T)$ which has a minimum at
$T\neq 0$ will lead to the same physical picture.

Another interesting issue is the type 0 definition of the 
scale-dependent Yang-Mills coupling. The approach adopted in
\JM\ and in this paper is to define 
the gauge coupling using  the static quark-antiquark potential 
computed from the Wilson loop  
as in  \Malda.
This is  a standard   definition  of the coupling used  in QCD 
and known as the $\V$-scheme (see, e.g.,  \sc\ and 
references therein).
One may compare  this definition 
with  the expression 
for the effective gauge coupling 
that follows from the  structure of the D3-brane action. 
 Careful consideration of the
D-brane effective action shows that in type 0 theory 
the effective 
Yang-Mills coupling is determined not just by the dilaton alone, but
{\it also}  by the tachyon field.
 Indeed, due to the existence of
a tachyon tadpole on the D-brane, the Born-Infeld part of the 
effective action of an electric D3-brane is  of
the form (here we set $2\pi \a'=1$) 
\eqn\coupnew{
\int d^4 x \ k(T)\   e^{-\Phi} \sqrt{-\det( G_{\a\b} +
G_{ij} \del_\a X^i  \del_\b X^j +  F_{\a\b} )}\ , }
where $k(T)= 1 + \four T + O(T^2)$.
The coefficient $1/4$ of the tachyon tadpole was found in \KT.
The reparametrization invariance and $T$-duality 
imply that the function $k(T)$ 
multiplies $\sqrt {-\det (G + F)}$ in the BI 
part of the action.\foot{This form of the action was independently
proposed in a recent paper \GA. Both we and Garousi \GA\ have checked the 
$T F_{\a\b} F^{\a\b}$ coupling that follows from this
action by calculating the corresponding 
3-point function on a disk. Ref. \GA\ also contains  
some further checks of the consistency of this form of the action. }
From the function that multiplies $F^2_{\a\b}$ in the D3-brane action
we read off that  
\eqn\yang{
{1\over \gym}\  \sim \  \ k(T) \  e^{-\Phi}  \ . } 
With  this  definition of the coupling (`the D-scheme')
the tachyon field plays a role in determining
the $u$-dependence of the Yang-Mills coupling
 (in the UV  one has 
$T\approx -1+ {4\ov \ln u}$). If we retain only the linear term in
$k(T)$ then we find, however, that 
 the  product on the r.h.s. of \yang\
grows as $\ln^2 u$ in the UV which disagrees with the $\V$-scheme.
One possible source of this disagreement is that
one needs to know the exact expressions for both $f(T)$ and
$k(T)$. If a zero of $k(T)$ coincides with the minimum of $f(T)$,
then we will find $\sim \ln u$ on the r.h.s. of \yang.\foot{
For example, with $f(T)=1+T+\ha T^2$ and $k(T)=1+T$,\  \ \ 
$k(T) e^{-\Phi}$ grows as $\ln u$, i.e.
 gives $  \ln u  -  { 39 \ov 8 } \ln \ln u  + ... $,
which is the same answer as the effective potential \neweffpot.}
It is also possible that $u$ is not simply proportional to the energy
scale of the gauge theory and that $\ln^2 u$ is actually to be
identified with $\ln (L_0/L)$. More work is needed to resolve these
issues.

\newsec{Acknowledgements}
We are grateful to J. Minahan,
A.M. Polyakov and E. Witten for very useful discussions.
The work  of I.R.K. was supported in part by the NSF
grant PHY-9802484 and by the James S. McDonnell
Foundation Grant No. 91-48.
The  work  of A.A.T. was supported in part
by PPARC, the European
Commission TMR programme grant ERBFMRX-CT96-0045
and  the INTAS grant No.96-538.

\vfill\eject
\listrefs
\end